\documentclass[acmsmall]{acmart}
\usepackage{multirow}
\usepackage{amsmath}
\usepackage{makecell} 
\usepackage{hhline} 
\usepackage{tabularx} 
\usepackage{xcolor}
\usepackage{hyperref}
\usepackage{url} 
\usepackage{amsthm} 
\usepackage{enumitem}

\AtBeginDocument{%
  }

\setcopyright{none}
\acmDOI{XXXXXXX.XXXXXXX}

\begin{document}

\title[Omission of \textit{Global South} from AI Music Generation]{Missing Melodies: AI Music Generation and its "Nearly" Complete Omission of the \textit{Global South}}

\author{Atharva Mehta}
\authornote{Both authors contributed equally to this research.}
\email{atharva.mehta@mbzuai.ac.ae}
\orcid{1234-5678-9012}
\affiliation{%
  \institution{Research Assoc. at Mohamed bin Zayed University of Artificial Intelligence}
  \state{Abu Dhabi}
  \country{UAE}
}

\author{Shivam Chauhan}
\authornotemark[1]
\email{shivam.chauhan@mbzuai.ac.ae}
\affiliation{%
  \institution{Research Assoc. at Mohamed bin Zayed University of Artificial Intelligence}
  \state{Abu Dhabi}
  \country{UAE}
}

\author{Monojit Choudhury}
\email{monojit.choudhury@mbzuai.ac.ae}
\affiliation{%
  \institution{Professor at Mohamed bin Zayed University of Artificial Intelligence}
  \state{Abu Dhabi}
  \country{UAE}
}








\renewcommand{\shortauthors}{Mehta et al.}

\begin{abstract}
Recent advances in generative AI have sparked renewed interest and expanded possibilities for music generation. However, the performance and versatility of these systems across musical genres are heavily influenced by the availability of training data. We conducted an extensive analysis of over one million hours of audio datasets used in AI music generation research, and manually reviewed more than 200 papers from eleven prominent AI and music conferences and organizations 
to identify a critical gap in the fair representation and inclusion of the musical genres of the \textit{Global South} in AI research.

Our findings reveal a stark imbalance: approximately 86\% of the total dataset hours and over 93\% of researchers focus primarily on music from the \textit{Global North}. However, around 40\% of these datasets include some form of non-Western music, and genres from the \textit{Global South} account for only 14.6\% of the data. Furthermore, approximately 51\% of the papers surveyed concentrate on symbolic music generation, a method that often fails to capture the cultural nuances inherent in music from regions such as South Asia, the Middle East, and Africa.
As AI increasingly shapes the creation and dissemination of music, the significant underrepresentation of music genres in datasets and research presents a serious threat to global musical diversity, including evaluating music generation models, economic disparities, and reinforcing biases in generative music models. We also propose some important steps to mitigate these risks and foster a more inclusive future for AI-driven music generation. 
\end{abstract}

\ccsdesc[500]{Applied computing~Sound and music computing}
\ccsdesc[300]{Information systems~Multimedia content creation}
\ccsdesc[300]{Computing methodologies~Artificial intelligence}
\ccsdesc[100]{General and reference~Surveys and overviews}

\keywords{Global South, AI Music Generation, Music Genre, Dataset, Deep Learning, Machine Learning}


\maketitle

\section{Introduction}
\label{sec:introduction}

Music has always been a crucial element in representing the traditions of different communities worldwide \cite{savage2019cultural}. With recent developments in AI, particularly through the use of deep learning models \cite{copet2024simple,schneider2024mousai,melechovsky-etal-2024-mustango,agostinelli2023musiclm}, there has been a dramatic improvement in generating music automatically. These advancements have led to the creation of various AI-driven music platforms, such as Jukebox \cite{dhariwal2020jukebox}, \hyperlink{https://suno.com/}{Suno}, \hyperlink{https://www.udio.com/}{Udio}, \hyperlink{https://boomy.com/}{Boomy}, \hyperlink{https://soundraw.io/}{Soundraw}, etc, giving users the ability to create music based on their preferences using text prompts. In terms of generation quality and performance, AI systems have significantly improved on automatic evaluation metrics as well as human evaluations, enabling more human-like and controlled music generation, indicating a significant improvement in their capabilities~\cite{melechovsky-etal-2024-mustango, agostinelli2023musiclm, schneider2024mousai, copet2024simple}. Due to the ease of use, people can now "vibe compose" musical pieces with minimal musical knowledge. \citet{article2018} noted, "AI-generated music could easily surpass the amount of music that has ever been recorded, due to how quickly it can be produced." 
In terms of its consumption, Mubert, an online AI music composition website, reported that by mid-2023 its users had streamed more than 100 million AI-generated tracks\footnote{\url{https://mubert.com/blog/an-in-depth-study-into-ai-music-its-creators-composers-and-adopters}}.
With growing compute and efficient models, systems like Mousai can generate 43-second music clips in 49.2 seconds using just 5.04 GB GPU memory, enabling round-the-clock production far beyond human efficiency.
\begin{figure*}[t!]
\centering
\includegraphics[width=0.91\textwidth, height=0.46\textwidth, trim={0 0.6cm 0 0},clip]{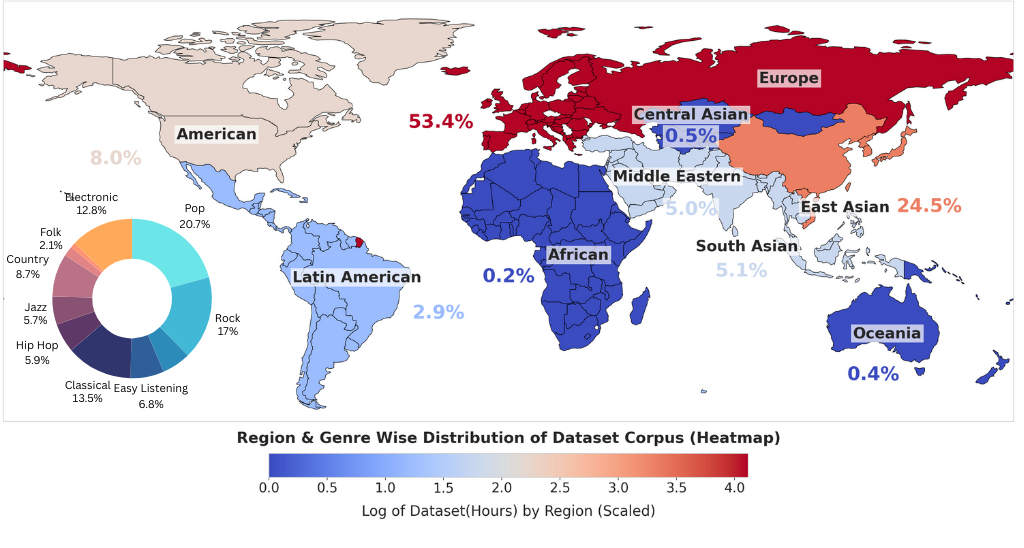}
\caption{%
Global Divide in AI Music Datasets: The heatmap shows the stark imbalance in regional musical styles (on the map) and genre (pie chart) representation in popular music datasets. 
}
\label{fig:dataset_infographic}
\end{figure*}

As noted in past studies \citep{10.1093/pnasnexus/pgae346, 10.1145/3640457.3688102}, despite impressive technological progress, biases remain prevalent in AI music generation systems. These include popularity, training data bias, and limited research interest, which leads to under-representation of musical genres from the \textit{Global South} regions of the world. For instance, the model described in \citet{melechovsky-etal-2024-mustango, copet2024simple, schneider2024mousai} often defaults to Western tonal and rhythmic structures when tasked with generating non-Western music such as Indian classical or traditional Middle Eastern genres, let alone lesser-known styles and genres such as the Baul (See \ref{sec:Definitions}) and instruments such as the Gonje (definition in \ref{sec:Definitions}). As a result, a generated piece intended to mimic an Indian raga may sound like a Western pop melody played on a sitar. 
Similarly, SunoAI, when attempting to generate  Maqamat (see \ref{sec:Definitions}) music of the Middle East, may round off the microtones to the nearest Western equivalent, resulting in a piece that lacks the distinctive sound of Arabic music.

As generative models continue to gain traction in the field of music generation, the misrepresentation and under-representation of the musical genres of the ``global majority" poses a significant threat to the inclusion of musical genres from around the world. The skewed distribution in datasets, reflected in model outputs, can lead to several issues, including cultural homogenization, reinforcement of Western culture dominance\footnote{\url{https://www.nytimes.com/2016/06/26/opinion/sunday/artificial-intelligences-white-guy-problem.html}}, misrepresentation of musical styles, and most importantly, gradual decline leading to the endangerment and disappearance of many musical genres~\cite{tan-2021,decolon_pop,intercontinental2023ai}. Drawing inspiration from studies such as \citet{joshi2020state} and \citet{bender2018data}, which systematically analyze the under-representation of languages of the global majority in the field of natural language processing, we attempt to analyze the under-representation of musical genres in the field of AI music generation. 

In this paper, we examine the distribution of musical genres and regional music from around the world in the datasets that are used to train models and conduct research in AI-music. We also study and contrast the distribution of participation of researchers from different parts of the world to the latter. This allows us to examine whether regions underrepresented in datasets are also underrepresented in authorship, further reinforcing existing imbalances. To this end, we analyze many popular datasets and papers from eleven AI conferences and organizations and contrast these findings with the digital (online) of consumption \& availability and ideal (offline) proxies of population. Our study reveals almost a complete omission of the musical styles of the \textit{Global South} in the AI world (Figure~\ref{fig:dataset_infographic} and Table~\ref{tab:genre_region_table}). Approximately 86\% of the total hours in available datasets and 93\% of papers focus majorly on the music of the \textit{Global North}. In contrast, only 14.6\% of the total hours of music data and 6.1\% of papers are first-authored by researchers from South Asian, Middle Eastern, Oceania, Central Asian, Latin American, and African countries.
This distributional skew in the training data is reflected in the quality of generated samples from several popular systems, Jukebox, Suno, and Udio (some examples are available for listening on our GitHub page\footnote{\href{https://atharva20038.github.io/aimusicexamples.github.io/}{https://atharva20038.github.io/aimusicexamples.github.io/}.}). 
We discuss several implications of this representational skew on the resulting quality of the generated music, research, and society, including concerns ranging from improper evaluations of generative models and limited creativity and diversity in generative AI-assisted music composition for the underrepresented genres to cultural erosion.

The article is organized as follows: In Section~\ref{sec:methodology}, we formally define some key concepts and present our research questions. We also explain in detail our data collection and analysis pipelines. Findings are presented in Section \ref{sec:results}, and their broader implications and mitigation strategies are presented in Section \ref{sec:discussion}. We conclude in Section \ref{sec:conclusion}. 

\section{Research Methods and Materials} 
\label{sec:methodology}

\subsection{Definitions \& Research Questions} \label{rqs}
We begin by mathematically defining fairness and representation. 
\begin{definition}
[\textbf{Representation}] 
Suppose $G = \{g_1, g_2, \dots\}$ is the set of genres (example: `classical', `folk') and $R = \{r_1, r_2, \dots\}$ the set of regions (example: `Oceania', `Latin America'). 
We define a musical {\em style}, $s = \langle g,r \rangle \in G \times R$, as a combination of a basic genre of music with a region, e.g., Latin American jazz music or folk music from Oceania.
\begin{equation*}
\begin{minipage}{0.48\textwidth}
\[
\scriptsize\textbf{Dataset-based:} \quad
\delta_{g} = \frac{D_{g}}{\sum\limits_{g ' \in G} D_{g'}}, \quad
\delta_{r} = \frac{D_{r}}{\sum\limits_{r' \in R} D_{r'}}
\]
\end{minipage}
\hfill
\begin{minipage}{0.48\textwidth}
\[
\scriptsize \textbf{Publication-based:} \quad
\rho_{s} = \frac{P_{s}}{\sum\limits_{s' \in G \times R} P_{s'}}, \quad
\alpha_{r} = \frac{P_{r}}{\sum\limits_{r' \in R} P_{r'}}
\]
\end{minipage}
\end{equation*}
\end{definition}
Here, $D_g$ and $D_r$ represent the dataset sizes in hours for a specific genre $g$ or region $r$, respectively; $P_s$ is the number of surveyed papers focusing on style $s$; $P_r$ is the number of surveyed papers with a first author from region $r$.

The {\em performance} of a model in a musical style is proportional to the amount of music data from that style that was used to train the model. This seemingly self-evident fact has also been empirically confirmed in numerous studies in machine learning literature. See~\citet{henning-etal-2023-survey}, for example, for a survey of such studies for language data and resultant performance disparity in LLMs. It follows that fair or equitable performance of models across musical styles implies an equivalently fair or equitable representation of the style (i.e., $D_{g}$ or $D_{r}$) in the dataset. However, there is no single or universal notion of {\em fairness}. For our purposes, we adopt the Rawlsian notion of fairness \cite{rawls1958justice} defined as the principle of least difference. More precisely,
\begin{definition}
[\textbf{Fairness}]
A music generation system is said to be fair across styles, if the difference between the quality of generated pieces in different styles is minimum, which is to say that the quantities $\max_{g \in G}\delta_{g} - \min_{g \in G} \delta_{g}$ and $\max_{r \in R}\delta_{r} - \min_{r \in R} \delta_{r}$ are minimized.
\end{definition}
Similarly, fairness of representation could also be defined with respect to publication-based representation of styles and regions, $\rho_s$ and $\alpha_r$. 
Fairness can also be defined from a utilitarian perspective, arguing for proportionate representation of styles in datasets and publications to their real-world utility. Although it is difficult to estimate the ``utility" of a musical style, one could approximate it through the real-world production and consumption patterns. To this end, we propose two {\em proxy of the real-world utility} of musical styles, namely, availability and consumption patterns in the digital space and the population of the corresponding region. 
\begin{definition}
[\textbf{Digital and Ideal Proxies of Utility}] 
We define digital proxies, $\pi_{dl}$, as a collection of digital music listenership and $\pi_{da}$, as the digital availability of styles and regions. Similarly, we define the ideal proxy, $\pi_{id}$, as the regional population : $\langle \pi_{pp} \rangle$.
\end{definition}
A fair representation in a utilitarian sense would imply a high positive correlation between the digital and ideal proxies of a style with its data-based and publication-based representations.
We call population the {\em ideal} proxy because the digital space is also an unfair or skewed representation of the real world, and ideally, the number of people interested in consuming and producing a particular style of music should be considered when estimating the true utility of a style. We must also recognize the limitations of {\em population} as a proxy for real-world consumption or popularity of a musical style, especially because several genres and traditions of music coexist among all populations, and their relative prevalence must be factored in. Nevertheless, due to the lack of better proxies, the population has been commonly used as an important indicator for studying disparities in the digital and technological world \cite{florida-2010, joshi2020state}. 
Building on these definitions, we formulate the following research questions: 
\begin{enumerate}[label=\textbf{RQ\arabic*}]
    \item What is the extent and fairness of the representation of various genres $(\delta_g)$ and regional music $(\delta_r)$ in the datasets?
    \item What is the extent and fairness of the representation of styles and regions in the research as captured by $(\alpha_r)$ and $(\rho_s)$?
    \item What is the correlation between the real-world proxies ($\pi_{dl}$, $\pi_{da}$ and $\pi_{id}$) and the data and publication-based representations?
    \item What are the implications of skewed representations in AI-music research on the global music ecosystem?
\end{enumerate}

\subsection{Data Collection}
To get our initial pool of papers, we employed a multi-stage, keyword-based paper selection method, leveraging the Scholarly package \cite{cholewiak2021scholarly} to gather approximately 5000 papers. This included up to 1000 papers per query, using broad search terms such as “music,” “music generation,” “non-western music,” “MIDI,” and “symbolic music.” We then refined our selection by focusing on papers presented at eleven major conferences, including \textit{AAAI, ACM, EUSIPCO, EURASIP, ICASSP, ICML, IJCAI, ISMIR, NeurIPS, NIME, and SMC}, chosen based on their popularity and prestige, narrowing our pool to around 800 papers.

\textbf{Dataset Papers}
To identify papers proposing datasets, we read through the title and abstract of each paper. This led to a set of 152 papers\footnote{\url{https://github.com/atharva20038/aimusicexamples.github.io/blob/master/Surveyed\%20Papers/Dataset-Papers.md}} proposing new datasets with a total of over one million hours of music. These datasets were manually annotated for the region and genres covered, total hours of music data, and whether the dataset was annotated with other details (such as instruments, genre, and style). Papers that directly provided details of the distribution of data points across genres and regions were analyzed with the already available statistics. Unfortunately, several datasets did not offer substantial details necessary for our study. If a dataset contained over 10,000 hours of audio, we analyzed the metadata for each file to extract genre and regional tags which were then mapped to one of the following genre categories ($G$) -- Folk, Electronic, Pop, Classical, Rock, Easy Listening, Hip-Hop, Jazz, Country, Reggae \& Soul, Experimental and Others. The region tags were mapped to our predefined region categories ($R$) -- Europe, South Asia, East Asia, America, Africa, Latin America, Oceania, Middle East, and Central Asia(see Figure \ref{fig:dataset_infographic}). However, when the genre and region were not explicitly mentioned in either the paper or the metadata, we did not make any assumptions; thus, 7.9\% of the datasets totaling 5,772 hours were excluded from our analysis. The remaining annotated data in hours for each genre ($D_{g}$) and region ($D_{r}$) are used for further analysis.

\textbf{Research Papers}
We conducted another keyword-based filtering on the titles and abstracts of the papers (with keywords such as "generative AI", "music generation", "auto-encoders", and "transformers") to further restrict the pool to those that propose generative AI techniques for music generation. This led us to a pool of 244 papers\footnote{\url{https://github.com/atharva20038/aimusicexamples.github.io/blob/master/Surveyed\%20Papers/Music-Papers.md}} that were further annotated for genres using the same method used to annotate datasets and further form our set of published papers with style annotations ($P_{s}$). Due to the lack of accurate data on region-wise information on genres such as hours of data on Kenyan folk music, which would often be mapped to "folk" in most papers, we simply take the summation over all regions to represent $P_s$; i.e, $P_s = \sum_{r \in R}P_s$.
For regional representation, we focused on the first author’s institutional affiliation, based on the location of the institution rather than the geographical focus of the study in the paper ($P_r$). The institution's country for the first author was mapped to $R$. This analysis focuses on first authors, as they are often regarded as the lead contributors who take primary responsibility for the research and dedicate the most time and effort to its development. For instance, if the first author was affiliated with a university in France, the paper was labeled as European, ensuring a clearer picture of region-wise research representation.

\textbf{Proxy Estimation}
\label{sec:proxies}
For the digital music, we draw on data from \hyperlink{https://soundcharts.com/}{\textit{SoundCharts}} and the \hyperlink{https://musicbrainz.org/}\textit{MusicBrainz} API. In \textit{SoundCharts}, we examine the YouTube views of the top 50 most-viewed songs from each region, summing the total views to evaluate music listenership across regions. While the portal lacks data from \textit{Australia}, and the \textit{Middle East}, and does not provide a detailed breakdown within \textit{Asia}, it offers a rough comparison between the music from \textit{Global North} and \textit{South}. The \textit{MusicBrainz} API, on the other hand, codes the region where a song was recorded, offering an estimate of the number of songs produced in each region that exist in the digital space. For the ideal space, we use population as a proxy and we collect its data from Wikipedia\footnote{\url{https://en.wikipedia.org/wiki/List_of_continents_and_continental_subregions_by_population}}.

\section{Findings} \label{sec:results}

In this section, we present the findings of our analysis of datasets and research papers. 

\subsection{RQ1: Representativeness of the Datasets}


\subsubsection{\textbf{Genre-wise Distribution ($\delta_g$):}}
As shown in Figure \ref{fig:dataset_infographic}, \textit{Pop} music has the highest (20.7\%) representation followed by \textit{Rock} (17\%) and \textit{Classical} (13.5\%) genres. \textit{Country, hip-hop}\textit{, blues} and \textit{jazz} have a moderate (more than 5\%) representation while \textit{Folk} and \textit{experimental} music have a low representation of 2.1\%. The other genres have minimal to no representation ($\leq 1\%$), which includes music for Children, \textit{Indie-music}, and region-specific genres.

Table \ref{tab:genre_region_table} shows that \textit{Pop} music forms 200K+ hours of the collection while \textit{Folk} music constitutes only 20K hours of the collection. \textit{Pop, Rock, Classical} and \textit{Electronic} music genres each have more than 10\% representation and more than 100K hours of data available.

\begin{figure*}[t!]
\centering
\includegraphics[width=0.85\textwidth, height=0.54\textwidth]{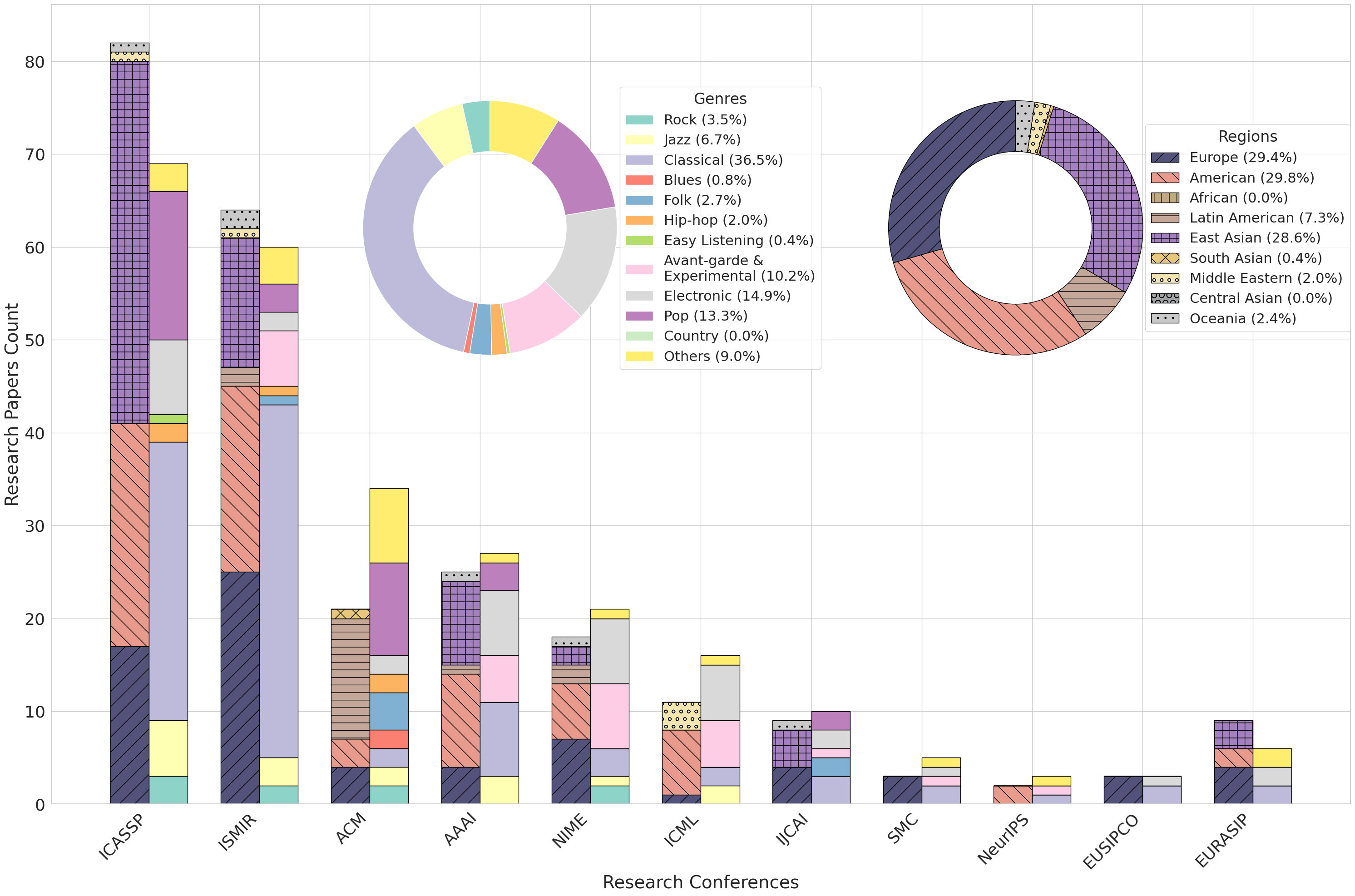}
\caption{%
Genre and Regional Representation ($\alpha_r/\rho_s$) in AI Music Research by Conferences: The bar plot illustrates the conference-wise representation of genre and region (derived respectively from the music datasets and the first author’s institutional affiliation). The pie charts show the overall representation by genre (left) and region (right).}
\label{fig:paper_infographic}
\end{figure*}
\begin{table*}[!t]
\centering
\scriptsize
\begin{minipage}[t]{0.48\textwidth}
\centering
\begin{tabular}{lrrrr}
\toprule
\textbf{Genre} & \textbf{MB} & \textbf{SC} & \textbf{Papers} & \textbf{Dur.} \\
\midrule
Pop & 49955 & \textbf{168.32} & 34 & \textbf{228.26} \\
Rock & \textbf{70823} & 106.89 & 9 & 186.67 \\
Electronic & 14025 & 115.73 & 38 & 140.43 \\
Classical & 8876 & 13.72 & \textbf{94} & 148.89 \\
Country & 2906 & 31.70 & -- & 95.77 \\
Hip-hop & 22562 & 87.87 & 5 & 64.38 \\
Jazz & 5663 & 11.38 & 17 & 62.20 \\
Blues & 13979 & 78.12 & 2 & 64.06 \\
Easy Listening & 486 & 1.00 & 2 & 74.41 \\
Folk & 9390 & 52.75 & 7 & 22.82 \\
Avant-garde/Exp. & 2655 & -- & 26 & 11.32 \\
Others & 9295 & -- & 23 & 0.94 \\
\bottomrule
\end{tabular}
\caption*{(a) Genre-wise distribution}
\end{minipage}%
\hfill
\begin{minipage}[t]{0.48\textwidth}
\centering
\begin{tabular}{lrrrrr}
\toprule
\textbf{Region} & \textbf{Pop} & \textbf{MB} & \textbf{SC} & \textbf{Papers} & \textbf{Dur.} \\
\midrule
European & 0.75 & 14433 & 2.38 & 70 & \textbf{6127.92} \\
East Asian & 1.66 & 662 & 58.39 & \textbf{84} & 2817.73 \\
South Asian & \textbf{2.07} & 671 & -- & 2 & 588.78 \\
Central Asian & 0.08 & 945 & -- & 0 & 57.01 \\
American & 0.58 & \textbf{28823} & \textbf{149.24} & 75 & 921.84 \\
Latin American & 0.66 & 2265 & 82.54 & 5 & 332.86 \\
Oceania & 0.05 & 484 & -- & 3 & 41.99 \\
African & 1.22 & 945 & 4.49 & 0 & 27.50 \\
Middle Eastern & 0.47 & 1366 & -- & 5 & 569.86 \\
\bottomrule
\end{tabular}
\caption*{(b) Region-wise distribution}
\end{minipage}
\caption{Distribution of datasets across (a) genres and (b) regions. Metrics: Duration (Dur., $10^3$ hours in (a), hours in (b)), MusicBrainz (MB, songs), SoundCharts (SC, billions), papers, and population (billions, regions only).}
\label{tab:genre_region_table}
\end{table*}

\subsubsection{\textbf{Region-wise Distribution ($\delta_r$):}}
We find that more than 6k hours of music in the research belongs to \textit{European} music and only 28 hours of music belong to \textit{African} music, as shown in Table \ref{tab:genre_region_table}. \textit{European}, \textit{East Asian} and\textit{ American} music are well represented in the datasets,  cumulatively having 85.9\% representation. On the other hand, \textit{South Asian and Middle Eastern} music has low representation, approximately 5\% each. \textit{Central Asian} and \textit{African} music have less than 1\% representation as depicted in Figure \ref{fig:dataset_infographic} showing they have extremely low representation. Notably, music research datasets often emphasize European classical music, which are often copyright-free, while US pop, country, and rock receive less attention, contributing to lower US participation in dataset creation compared to Europe.  As can be seen in the figure, a striking picture emerges when we plot the heatmap; \textit{Global South} is marked in blue, while \textit{Global North} appears in red, highlighting the imbalance in representation of regions in music research datasets.


\vspace{-0.5em}
\subsection{RQ2: Representativeness of Research}
\textbf{Genre-wise Distribution ($\rho_s$):} Figure \ref{fig:paper_infographic} shows that \textit{Classical} music is the most prominent genre in music generation research, accounting for 36.5\% representation in research. \textit{Electronic} music, with a substantial 14.9\%, and \textit{Pop}, at 13.3\%, follow as the next most studied genres. These genres likely benefit from a high volume of accessible data and established digital formats, making them more conducive to computational research. \textit{Avant-garde \& Experimental} genres have significant representation as well (10.2\%), underscoring interest in exploring unconventional and innovative sounds within AI-driven music generation.
Conferences such as \textit{ICASSP} and \textit{ISMIR} have a strong representation of \textit{Classical}, \textit{Pop}, and \textit{Electronic} genres, while \textit{NIME} and \textit{AAAI} showcase a diverse genre distribution, including \textit{Avant-garde \& Experimental} genres, indicating these venues' openness to exploring other genres also in music generation. Genres like \textit{Jazz} (6.7\%), \textit{Rock} (3.5\%), and \textit{Folk} (2.7\%) have moderate representation, suggesting a niche but valuable exploration of these genres in AI music research. \textit{Country} and \textit{Easy Listening} are among the most underrepresented genres in music generation research.

\textbf{Region-wise Distribution ($\alpha_r$):}
Figure \ref{fig:paper_infographic} highlights significant disparities in research focus between genres of music and affiliations of researchers, underscoring the under-representation of the \textit{Global South}. Cumulatively, representation of all regions in the \textit{Global North} is approximately 88\%, with Europe having the highest representation. In contrast,  \textit{Global South} has only about 12\% representation. This severe under-representation of \textit{Global South} researchers threatens the preservation of cultural heritage from regions such as \textit{Africa}, \textit{Middle East}, and \textit{South Asia}, which collectively account for less than 2.4\% representation.

Table \ref{tab:genre_region_table} shows that \textit{East Asia} (84 papers), \textit{Europe} (70 papers), and \textit{America} (75 papers) are leading in terms of researcher-affiliations, which is likely due to the strong music industries in these regions. However, regions like \textit{Africa}, \textit{Latin America}, and \textit{Oceania} show minimal or no representation.
At the ACM conferences including \hyperlink{https://www.acmmmasia.org/}{MMAsia}, \hyperlink{https://www.icmr2024.org/}{ICMR}, \hyperlink{https://www.aimlsystems.org/}{AIMLSystems}, 
\hyperlink{https://www.um.org/umap2024/}{UMAP}, \hyperlink{https://cc.acm.org/2022/}{C\&C}, \hyperlink{https://dl.acm.org/journal/TKDD}{TKDD}, \hyperlink{https://dl.acm.org/journal/tomm}{TOMM}
There is a notable emphasis on \textit{Pop}, with the majority of researchers affiliated with institutions coming from \textit{East Asia}. We found only one paper at ACM that had a first author from a \textit{South Asian} region.

\subsection{RQ3: Correlation between Representation and Utility}


Upon analyzing the digital and real-world proxies, the data reveals significant disparities across regions: \textit{American} music leads digital consumption with 150 billion listeners, yet \textit{European} music, with only 2.38 billion listeners, dominates research datasets (6,127.92 hours). Meanwhile, \textit{South Asia}, with the largest population of 2.07 billion, is severely underrepresented in research (588.78 hours) and digital catalogs (671 songs). Similarly, \textit{Africa}, despite a population of 1.22 billion, has minimal representation (27.50 hours) and digital catalogs (945 songs). \textit{East Asian} music shows moderate representation in research (2,817.73 hours), listenership (part of 58.9 billion listeners), and population (1.66 billion), but lags in digital catalogs (662 songs). 
The representation ratio of \textit{Global North} to \textit{Global South} music, as reflected in \textit{MusicBrainz}, reveals a stark contrast. In the digital realm, this ratio stands at 6.58, while in the ideal space, it drops significantly to 0.66. Additionally, the correlation between the representation of regions in datasets ($\delta_r$) and digital proxies is 0.35 ($Corr(\pi_{da}, \delta_r)$) and -0.16 ($Corr(\pi_{dl}, \delta_r)$), while the correlation between the publication based representation ($\alpha_r$) for regions and \textit{MusicBrainz} is 0.65 ($Corr(\pi_{da}, \alpha_r)$). This suggests a mild alignment between digital music distribution and the research space but highlights a disconnect between regional viewership and the representation of those regions in music datasets.
Based on this analysis, we assert that while viewership for \textit{European} music and its population is lower compared to \textit{Asian} and \textit{African} regions, \textit{Global North} music datasets disproportionately dominate the research space. In contrast, \textit{South Asian} and \textit{African} music remain significantly underrepresented in music datasets, even though their digital consumption and real-world presence are comparable to, and in some cases exceed, those of \textit{European} and \textit{American} regions.  

When compared to the digital space, we observe an interesting contrast with \textit{Hip-Hop} being highly represented with 22562 songs in the \textit{MusicBrainz} database, while having only 6\% representation, while \textit{Electronic} music has only 14025 songs but has 13.1\% representation. Using \textit{SoundChart} data we observed that \textit{Classical} music has only 13.72 billion views but has 13.5\% representation; meanwhile, \textit{Folk, Blues}, and \textit{Hip-Hop} have more than 50 billion views each but have 6\% or less representation.
Among styles represented in the datasets ($\rho_s$), pop is 3 times more popular than \textit{Hip-Hop}, but in the digital music market, the situation is just the reverse of this, where \textit{Hip-Hop} is 3 times more popular than \textit{Pop}. In contrast to the region-wise distribution, the research domain shows a strong correlation with the digital space in genre-specific scenarios, with correlation of 0.71 ($Corr(\pi_{da}, \rho_s)$) and 0.62 ($Corr(\pi_{dl}, \rho_s)$) for the genre representation relative to \textit{MusicBrainz} and \textit{SoundCharts} respectively. 
Genres like \textit{Rock}, \textit{Pop}, and \textit{Electronic} are well represented in digital space as well as in AI research. On the other hand, \textit{Country} and \textit{Classical}, while not popular in digital space, are well-represented in research. \textit{Hip-hop}, \textit{Experimental}, \textit{Folk}, \textit{Blues}, and \textit{Jazz} exhibit the opposite pattern; these genres are popular in the digital space, but not so in research. 

\section{Why Does it Matter?}\label{sec:discussion}

The marginalization of \textit{Global South} music genres in AI-driven music generation has profound implications. Historically, colonization has favored those who control technology while disadvantaging the colonized~\cite{technopoly}. As \citet{decolon_pop} notes, Western narratives continue to dominate global pop, with local traditions in Turkey, Brazil, and Peru gaining recognition only after Western reissues or collectors’ interest (e.g., Arşivplak, \hyperlink{https://www.djtahira.com/}{DJ Tahira}, \hyperlink{https://www.barbesbrooklyn.com/}{Barbès Records}). These cases highlight cultural homogenization, where value is conferred externally rather than internally sustained. Building on this concern, \citet{collins2024avoiding} warns that AI could intensify such dynamics by monopolizing music production, potentially erasing diverse traditions in favor of uniform outputs. As AI becomes further embedded in the music industry, the vibrant melodies of the Global Majority risk both erosion and extinction. This trajectory threatens not only the survival of these cultural treasures but also the richness of global musical diversity. In what follows, we examine the consequences of the underrepresentation of musical styles in AI-generated music and outline recommendations to mitigate its risks.

\begin{table*}[t!]
\centering
\scriptsize
\begin{tabularx}{\textwidth}{|X|X|X|X|l|}
\hline

\textbf{Category} & \textbf{Problem} & \textbf{Risks} & \textbf{Mitigation Strategy} & \textbf{Severity} \\ \hline
Moderately represented & Inconsistent quality and adherence to the prompt. & Produces music that may not match the prompt, impacting information retrieval. & Improve model architectures and use more descriptive prompts. & Moderate \\ \hline
Under-represented & Instruments and melodies are out of sync, resulting in poor-quality music. & Weak representation leads to less impactful music generation. & Use better datasets and warn users of potential inaccuracies. & High \\ \hline
Unrepresented & Music diverges entirely from the genre. & Misleads listeners, misrepresenting the genre digitally. & Avoid generating samples for unfamiliar genres. & Very High \\ \hline
\end{tabularx}
\caption{Problems of under-representation of musical genres in AI systems, their risks and mitigation strategies.}
\label{tab:suggest}
\end{table*}

{\bf Bias in Evaluation.}
Generative music models are often evaluated using audio embedding backbones such as PANN-CNN14 \cite{pann} and VGGish\footnote{\url{https://github.com/tensorflow/models/tree/master/research/audioset/vggish}}, which map audio into fixed-length embeddings for similarity metrics like Fréchet Audio Distance (FAD) \cite{kilgour2018fr} and KL divergence \cite{kullback1951information}. 
As we saw, global music datasets show strong regional and genre disparities, with several \textit{Global South} regions underrepresented. 
Large-scale backbone models, reliant on such datasets, are expected to inherit these gaps, limiting their ability to encode features such as timbres, rhythmic structures, and ornamentations. To better understand these biases, we compare the representation of guitar, piano, drums, globally popular instruments, to sitar, tabla, accordion, bagpipes regional instruments in the training data. In both VGGish and PANN-CNN14, 52.33\% and 67.54\% of the training clips contain at least one of {guitar, piano, drums}, whereas the union of {sitar, tabla, accordion, bagpipes} totals <3\%, and none of them individually has higher than 1\%.
Given these instruments’ central role in Hindustani Classical (sitar, tabla) and folk (accordion, bagpipes) traditions, their underrepresentation risks poor embedding quality and biased evaluations against such genres, akin to fairness issues observed in text embeddings \cite{nlp_women_bias}.

\textbf{Research and Technological Exclusion.} The exclusion of \textit{Global South} styles from AI training datasets restricts their evolution in the digital age, risking stagnation while others continue to expand. Despite rich traditions, researchers from the \textit{Global South} contribute only 12.1\% of publications (Section~\ref{sec:results}), with most work centered on East Asian, American, \& European music (Figure~\ref{fig:paper_infographic}).
As shown in Section~\ref{sec:results}, representation gaps vary: Oceania and Latin America show higher research representation than data representation, South Asia \& Middle East show the reverse, while Africa \& Central Asia remain critically low in both, risking loss of regional traditions. Similarly, for genres  Folk, Blues, Country, \& Easy Listening remain under 3\%, far behind Pop, Electronic, and Classical.

{\bf Dominance of Symbolic Music Generation.} As a result of these imbalances, symbolic music generation has become the dominant paradigm, accounting for 51\% of surveyed works and nearly 90\% of ACM publications. Its prevalence stems from its alignment with the structured conventions of \textit{Global North} genres. However, this approach is inadequate in capturing the improvisation, ornamentation, and microtonality that are central to many \textit{Global South} traditions (e.g., Ragas, poly-rhythms, Maqam). Continued infrastructure development centered on symbolic methods risks further marginalizing these traditions, excluding them from the benefits of future advancements. Further, if left unchecked, such biases risk cultural erosion and deepening economic divides \cite{okolo2023ai,weforum2023aidivide}.

\textbf{Cultural Erosion.}
By focusing primarily on \textit{Global North} music datasets, AI models risk reinforcing existing cultural biases and further marginalizing musicians and composers from the \textit{Global South}. This imbalance is evident in our analysis, where the ratio of \textit{Global North} to \textit{Global South} representation stands at 6.09 in research datasets and rises to 6.58 in the digital music space, with $Corr(\pi_{da}, \delta_g) = 0.71$, $Corr(\pi_{da}, \delta_r) = 0.35$, and $Corr(\pi_{da}, \alpha_r) = 0.65$ indicating a strong correlation between the digital space and the dataset/research publication space. As AI-generated music scales, it begins to feed back into the digital space, creating a self-reinforcing vicious loop.

\textbf{The Vicious Loop of Synthetic Music Data.}
Synthetic data generation can reinforce and amplify existing biases in music representation. The skew in representation in the larger data is often reflected in the synthetically generated music data. As noted by \citet{agarwal2021fairness}, training on model-generated data can create feedback loops that exacerbate disparities. To illustrate, the \textit{MusicBench} dataset \cite{melechovsky-etal-2024-mustango}, which augments \textit{MusicCaps} \cite{agostinelli2023musiclm}, expands from 5k to 37k samples while preserving its skewed genre distribution, boosting the absolute counts of overrepresented genres such as classical (13.7\%), electronic (15.6\%), and rock (10.5\%), and leaving Asian (2.6\%), hip-hop (3.4\%), and country (5.3\%) comparatively scarce. Such scaling risks locking models into biased generation patterns, further hindering fair musical diversity.

\section{Conclusion and Recommendation} \label{sec:conclusion}
In this article, we made an attempt to bring out an important gap in AI-driven music generation: the noticeable lack of representation of music genres from the \textit{Global South}. While AI tools have made incredible progress in generating music, they overwhelmingly focus on the musical styles from the \textit{Global North}, leaving out many unique and culturally rich styles from regions such as Africa, Latin America, South Asia, and the Middle East. Global musical cultures evolve through economic, technological, and community-driven efforts to decolonize and diversify~\cite{tan-2021}. In AI-generated music, fully egalitarian outcomes may be impossible~\cite{Choudhury_Deshpande_2021}, yet concrete strategies can still foster broader inclusion of diverse traditions. Here are some concrete steps that the AI-music research community can adopt to ensure that future AI-music generation systems can serve everyone on the planet more equitably. 

\textbf{Explicit Genre Disclosure and Model Limitations:}
Music generation papers should clearly state the genres used for training and evaluation, following the Bender Rule~\cite{bender2018data}, to avoid misinterpretations about model universality. They should also acknowledge limitations such as symbolic models’ inability to capture microtonal variations (e.g., shrutis (See Appendix \ref{sec:Definitions})) or irregular rhythms~\cite{shah2001visualization} in genres like Indian Classical or Maqamat music. Transparency in genre coverage and constraints enables targeted improvements.

\textbf{Avoid Generation When Uncertain:}
Models will inevitably encounter genres they are under-trained on or cannot generate confidently, since an ideal egalitarian approach is difficult to achieve practically. In such cases, issue warnings for underrepresented genres and avoid generating for unrepresented ones to prevent misleading outputs and distortion of the digital music space. Table \ref{tab:suggest} summarises the issues, risks, and our recommendations.

\textbf{Investing in Inclusive Datasets:} 
Initiatives like the Masakhane Project~\cite{orife2020masakhane} in NLP show how community-driven, open-source efforts can build representative datasets. A similar approach in music could ensure diverse regional and genre coverage, addressing gaps such as \textit{Hip-Hop}, which is digitally prominent but underrepresented in research. Following examples like Google’s Inclusive Images Competition~\cite{atwood2020inclusive}, AI music generation should balance datasets by both genre and region~\cite{bryan2024reducing}.

\textbf{Transfer Learning for Underrepresented Styles: } Similar to the language domain, where transfer learning is used for low-resource languages~\cite{transfer,zoph2016transfer,info14120638}, the music research community should invest in techniques for sample-efficient cross-genre transfer of musical styles, instruments, melodic structures, and so on \cite{doosti2023transferlearningunderrepresentedmusic}.

\textbf{Inclusive Evaluation: }
Genre-specific evaluations are rarely conducted or clearly reported, leaving the performance of many styles unknown. Prior work~\cite{xiong2023comprehensive,yang-2018} highlights the need for subjective assessments, while automatic evaluations require backbone models finetuned on diverse music to avoid misrepresentation (See Section \ref{sec:discussion}). Researchers should also develop metrics that generalize across genres without repeated finetuning.
    




\bibliographystyle{ACM-Reference-Format}
\bibliography{sample-base}

\appendix
\section{Tabular Data and Definitions}

\subsection{Definitions}
\label{sec:Definitions}
\begin{enumerate}
    \item The Bauls are mystic minstrels from the Bengal region spread across India and Bangladesh, blending Sufism and Vaishnavism in songs about the love between the human soul and a personal god within.
    \item The Gonje is a one-stringed West African fiddle, often made with a snakeskin-covered gourd and horsehair string, played solo or in ensembles with other traditional instruments.
    \item Maqam, in traditional Arabic music, is a melodic mode system defining pitches, patterns, and improvisation, central to Arabian art music, with 72 heptatonic scales.
    \item The Shruti, in Indian Classical, is the smallest interval of pitch that the human ear can detect and a singer or musical instrument can produce.
\end{enumerate}

\subsection{Instruments Data}
\label{sec:Instruments_Collection}
\begin{table}[!htbp]
\centering
\begin{tabular}{c c c}
    \toprule
    \textbf{Instrument} & \texttt{VGGish (\%)} & \texttt{PANN (\%)} \\
    \midrule
    \texttt{Guitar}     & \textbf{27.39} & \textbf{43.97} \\ 
    \texttt{Drums}      & 19.14 & 13.75 \\
    \texttt{Piano}      & 5.80  & 9.82 \\
    \midrule
    \texttt{Accordion}  & 1.77  & 2.5  \\ 
    \texttt{Clarinet}   & 0.24  & 1.81 \\
    \texttt{Bagpipes}   & 0.20  & 1.52 \\ 
    \texttt{Sitar}      & 0.09  & 1.35 \\ 
    \texttt{Tabla}      & 0.08  & 1.47 \\
    \texttt{Bassoon}    & 0.08  & --   \\
    \bottomrule
\end{tabular}
\caption{Percentage Composition of Instruments in the Collection for VGGish and PANN Models}
\end{table}
\end{document}